\documentclass[aps,prl,amsmath,amssymb,reprint,superscriptaddress,longbibliography]{revtex4-2}
\usepackage{xcolor}
\usepackage{graphicx}
\usepackage{dcolumn}
\usepackage{bm}
\usepackage{braket}
\usepackage{verbatim} 
\usepackage[english]{babel}
\usepackage{xfrac}
\usepackage{soul}
\usepackage{comment}
\usepackage{braket}
\usepackage[explicit]{titlesec}

\begin{document}


\title{Thermalization within a Stark manifold through Rydberg atom interactions}


\author{Sarah E. Spielman}
\affiliation{Department of Physics, Bryn Mawr College, Bryn Mawr, PA 19010.}

\author{Sage M. Thomas}
\affiliation{Department of Physics, Bryn Mawr College, Bryn Mawr, PA 19010.}

\author{Maja Teofilovska}
\affiliation{Department of Physics, Bryn Mawr College, Bryn Mawr, PA 19010.}

\author{
Annick C van Blerkom}
\affiliation{Department of Physics, Bryn Mawr College, Bryn Mawr, PA 19010.}

\author{Juniper J. Bauroth-Sherman}
\affiliation{Department of Physics and Astronomy, Ursinus College, Collegeville, PA 19426.}

\author{Nicolaus A. Chlanda}
\affiliation{Department of Physics and Astronomy, Ursinus College, Collegeville, PA 19426.}

\author{Hannah S. Conley}
\affiliation{Department of Physics and Astronomy, Ursinus College, Collegeville, PA 19426.}

\author{Philip A. Conte}
\affiliation{Department of Physics and Astronomy, Ursinus College, Collegeville, PA 19426.}

\author{Aidan D. Kirk}
\affiliation{Department of Physics and Astronomy, Ursinus College, Collegeville, PA 19426.}

\author{Thomas J. Carroll}
\affiliation{Department of Physics and Astronomy, Ursinus College, Collegeville, PA 19426.}

\author{Michael W. Noel}%
\affiliation{Department of Physics, Bryn Mawr College, Bryn Mawr, PA 19010.}




\date{\today}

\begin{abstract}
We use dynamical quantum typicality to predict the thermal equilibrium state of ultracold Rydberg atoms exchanging energy via long-range dipole-dipole interactions. We excite atoms to the center of a manifold of nearly-harmonically spaced clusters of Stark energy levels and then allow them to equilibrate. Comparing the equilibrium state to our thermal prediction across a range of Rydberg densities, we find that this system generally fails to thermalize, though it approaches the thermal state at the highest tested density. This is the first direct comparison of a dynamical quantum typicality calculation to experiment.
\end{abstract}


\maketitle

An open quantum system thermalizes through interactions with the environment. Remarkably, an isolated quantum system can also thermalize by acting as its own reservoir. The strong version of the eigenstate thermalization hypothesis (ETH) explains this by postulating that all eigenstates of the Hamiltonian are thermal~\cite{deutsch_QuantumStatisticalMechanics_1991,srednicki_ChaosQuantumThermalization_1994,srednicki_ThermalFluctuationsQuantized_1996,rigol_ThermalizationItsMechanism_2008}. In this picture, an out-of-equilibrium initial state amounts to a specially arranged superposition of the energy eigenstates with nonthermal expectation values. Thermalization then occurs because the components of the initial state dephase with time~\cite{nandkishore_ManyBodyLocalizationThermalization_2015}. However, exceptions to the ETH have been observed, such as many-body localization and quantum many-body scars. Therefore, a weak version of the ETH (wETH), which requires only the vast majority of eigenstates to be thermal, has been proposed~\cite{biroli_EffectRareFluctuations_2010,ikeda_FinitesizeScalingAnalysis_2013,beugeling_FinitesizeScalingEigenstate_2014,iyoda_FluctuationTheoremManyBody_2017,yoshizawa_NumericalLargeDeviation_2018}.

Closely related to the ETH is dynamical quantum typicality, which states that, for a large enough Hamiltonian, the vast majority of pure states with similar expectation values at the initial time will continue to have similar expectation values at all times~\cite{bartsch_DynamicalTypicalityQuantum_2009,sugiura_CanonicalThermalPure_2013,reimann_DynamicalTypicalityApproach_2018,reimann_DynamicalTypicalityIsolated_2018,beckemeyer_QuantumTypicalityApproach_2026,capizzi_HydrodynamicsEigenstateThermalization_2025}. If the Hamiltonian satisfies the wETH, then those expectation values will be the same as predicted by the microcanonical ensemble~\cite{ikeda_FinitesizeScalingAnalysis_2013,reimann_DynamicalTypicalityApproach_2018}. For sparse Hamiltonians in particular, quantum typicality can therefore efficiently calculate the thermal state of an isolated system~\cite{elsayed_RegressionRelationPure_2013,steinigeweg_PushingLimitsEigenstate_2014}.

The dynamics of thermalization in isolated systems have been studied experimentally, including trapped ion systems~\cite{clos_TimeResolvedObservationThermalization_2016}, superconducting qubits~\cite{chen_ObservationStrongWeak_2021,zhu_ObservationThermalizationInformation_2022}, and nitrogen-vacancy centers~\cite{martin_ControllingLocalThermalization_2023}. Systems that fail to thermalize have also been studied extensively in recent years. This includes many-body localization~\cite{luschen_ObservationSlowDynamics_2017}, Hilbert space fragmentation~\cite{sala_ErgodicityBreakingArising_2020}, and quantum many-body scars~\cite{turner_QuantumScarredEigenstates_2018,turner_WeakErgodicityBreaking_2018,evrard_QuantumScarsRegular_2024,evrard_ManyBodyOscillationsThermalization_2021,spielman_QuantumManybodyScars_2024}. Cold atoms, in particular, provide an ideal platform for studying equilibration and thermalization~\cite{ueda_QuantumEquilibrationThermalization_2020}. Thermalization was recently observed in a Rydberg atom array~\cite{zhao_ObservationQuantumThermalization_2025} and in Bose-Einstein condensates~\cite{kaufman_QuantumThermalizationEntanglement_2016,prufer_CondensationThermalizationEasyplane_2022}. The relaxation of the magnetization in a spin-1/2 model was observed using Rydberg atoms in an amorphous cloud~\cite{orioli_RelaxationIsolatedDipolarInteracting_2018}.

In this letter, we study the equilibrium state of an isolated system of cold Rydberg atoms, excited in a magneto-optical trap, which interact via few-body dipole-dipole interactions. We numerically predict the thermal state using dynamical quantum typicality and compare directly, for the first time, to experimental results at a wide range of densities. We find that while our system equilibrates to a steady state, it does not thermalize.

\textit{Experiment\textemdash}Our experimental platform consists of an amorphous sample of ultracold Rydberg atoms that change electronic states through resonant dipole-dipole interactions. We apply an electric field to Stark shift the electronic states, which group into clusters that form the nearly-equidistant ``rungs'' of an energy ladder.  A cluster near the center of this ladder is initially excited, after which long-range interactions redistribute energy to the clusters above and below until a steady state is reached.  Following a set interaction time, we quantify this redistribution using a combination of microwave spectroscopy and selective field ionization. We repeat this process for a broad range of Rydberg densities.

\begin{figure}
 \includegraphics{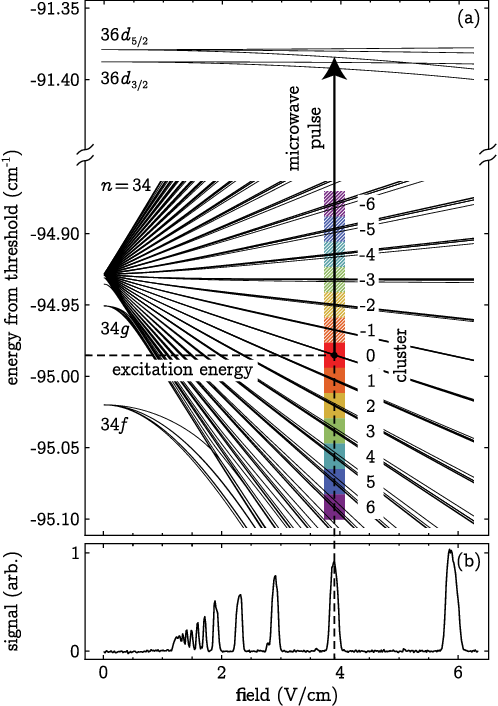}
\caption{\label{fig:starkmap}(a) Stark map showing the $|m_j|=1/2, 3/2$, and 5/2 states of the $n=34$ manifold and the $36d$ states. Manifold states are organized into clusters of energy levels with nearly-harmonic spacing of approximately 530 MHz. The initially excited manifold cluster is highlighted in red and labeled 0. During an interaction time of 3~$\mu$s and at a static field of 3.9~V/cm, resonant dipole-dipole interactions transfer population to clusters above and below the initial cluster, which are highlighted in different colors and labeled with negative and positive integers respectively. A microwave pulse is scanned over frequency to transfer the population of each manifold cluster to a $36d$ state, where it can be resolved with state selective field ionization. (b) An electric field scan at a particular wavelength of the Rydberg excitation laser, showing each manifold state intersected by the horizontal dashed line in (a).}
\end{figure}

We first cool and trap about $10^6$ rubidium-85 atoms in a magneto-optical trap using two 780~nm lasers that drive the $5s_{1/2} \rightarrow 5p_{3/2}$, $F=3 \rightarrow F=4$ cycling transition and the $F=2 \rightarrow F=3$ hyperfine pumping transition. To reach $n=34$ Rydberg states, pulsed 776~nm and 1265~nm lasers excite from $5p$ to $5d$ and then from $5d$ to Rydberg states with $nf$ character, respectively. The width of the Rydberg excitation pulse is 250~ns.

In Fig.~\ref{fig:starkmap}(a) we see how the Stark effect shifts the energy eigenstates with $|m_j| = \frac{1}{2}, \frac{3}{2},$ and $\frac{5}{2}$.  The initially degenerate high angular momentum states at $n=34$ fan out as the electric field increases, forming a nearly-harmonic ladder.  Each rung of this ladder is actually a closely spaced cluster of different $m_j$ states \cite{zimmerman_StarkStructureRydberg_1979}.  

We first lock the frequency of the 1265~nm laser slightly below the energy of the zero-field high-$\ell$ $n=34$ manifold, shown by the horizontal dashed line in Fig.~\ref{fig:starkmap}(a). Each peak in Fig.~\ref{fig:starkmap}(b) shows a Stark cluster tuning into resonance with the excitation laser as the electric field increases. We choose the Stark cluster at about 3.9~V/cm as our initial state, which is colored red and labeled 0 in Fig.~\ref{fig:starkmap}(a).  Note that our laser excitation does not resolve the $m_j$ structure within a cluster. At this field, which is held constant throughout the experiment, the energy spacings between neighboring clusters are
nearly equal and about 530 MHz. The anharmonicity in the spacing between clusters is smaller than the cluster widths, which range from 13 to 180 MHz. This nearly-harmonic splitting arises from the primarily linear Stark shifts; in alkali atoms quantum defects are negligible for states with $\ell \geq 4$~\cite{zimmerman_StarkStructureRydberg_1979}.

Upon excitation to this initial cluster, atoms can immediately change electronic states through resonant dipole-dipole interactions and subsequently populate clusters across the manifold. Following the initial excitation, we set a fixed interaction time of 3~$\mu$s, which is smaller than the lifetime of these states but long enough for the state distribution to reach equilibrium. To quantify energy transport across the manifold, we measure the population distribution among six clusters above and below the initial cluster. 

\begin{figure}
    \centering
    \includegraphics{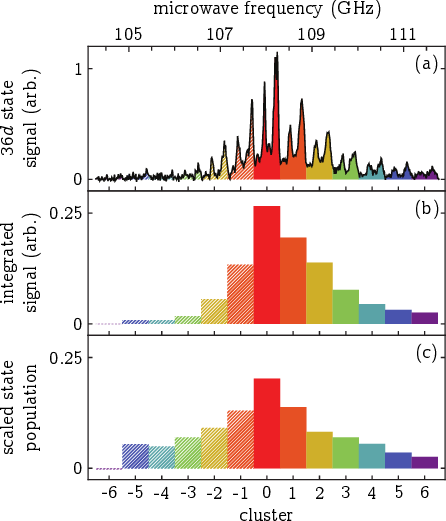}
    \caption{Example of data processing using a Rydberg atom density of about $10^{10}$~cm$^{-3}$. (a) The signal obtained from selective field ionization of the $36d$ states through microwave spectroscopy from the $n=34$ manifold, as a function of microwave frequency. The colored regions highlight the particular frequency ranges over which individual clusters in the manifold are driven to a $36d$ state, using the same coloring and labeling scheme as Fig.~\ref{fig:starkmap}(a). Each region has five visible peaks, which are associated with the coupling to the five $36d$ states. (b) The integrated signal within each of the colored regions shown in (a). (c) The integrated signal from (b) scaled according to the measured couplings between each cluster and the $36d$ states.}
    \label{fig:microwave-spectra}
\end{figure}

Selective field ionization alone cannot resolve the closely spaced Stark clusters. Instead, we use a 100~ns pulsed microwave field to drive population from the $n=34$ manifold to $36d$ states following the 3~$\mu$s interaction time. Standard selective field ionization can then resolve these $36d$ states from manifold states. We scan the microwave frequency over a range of 104 to 112~GHz and measure the resulting $36d$ state signal as shown by Fig.~\ref{fig:microwave-spectra}(a) for a Rydberg atom density of about $10^{10}$~cm$^{-3}$. One can imagine rotating Fig.~\ref{fig:microwave-spectra}(a) by 90 degrees clockwise and placing it on Fig.~\ref{fig:starkmap}(a), with the spectra in Fig.~\ref{fig:microwave-spectra}(a) resulting from scanning the microwave frequency across the 13 manifold clusters.

At 3.9~V/cm the $36d$ state splits into five states with $j=5/2$, $|m_j|=\frac{1}{2},\frac{3}{2},\frac{5}{2},$ and $j=\frac{3}{2}, |m_j|=\frac{1}{2},\frac{3}{2}$. The width of the $36d$ states is about 410 MHz, which is smaller than the energy spacing between neighboring manifold clusters, so we expect the microwave spectra of Fig.~\ref{fig:microwave-spectra}(a) to consist of five peaks within each Stark state cluster. We break the microwave spectrum into 13 frequency ranges, each associated with the transfer of a single cluster to the five $36d$ states.  These frequency regions are color coded to match the scheme presented in Fig.~\ref{fig:starkmap}(a). We integrate the signal in each of these 13 regions to find the total microwave signal associated with a given cluster as shown in Fig.~\ref{fig:microwave-spectra}(b).  

If the coupling strength between each cluster and the $36d$ states were uniform, Fig.~\ref{fig:microwave-spectra}(b) would measure the relative cluster populations.  In reality, the coupling strengths between each cluster and the $36d$ states vary considerably. In a separate experiment, using the technique described in~\cite{opsahl_EnergyTransportHighlyPolarized_2026}, we have measured these relative couplings yielding (0.005, 0.014, 0.014, 0.023, 0.060, 0.098, 0.125, 0.132, 0.159, 0.103, 0.080, 0.087, 0.087) for the state clusters labeled ($-6 \ldots 6$).  Dividing the integrated signal in  Fig.~\ref{fig:microwave-spectra}(b) by these coupling strengths and normalizing, we find the final population distribution shown in Fig.~\ref{fig:microwave-spectra}(c). Note that the coupling strength for the first few clusters (-6, -5, -4, and -3) is quite weak, leading to greater uncertainty on the negative side of the distribution.

\begin{figure}[t!]
 \includegraphics{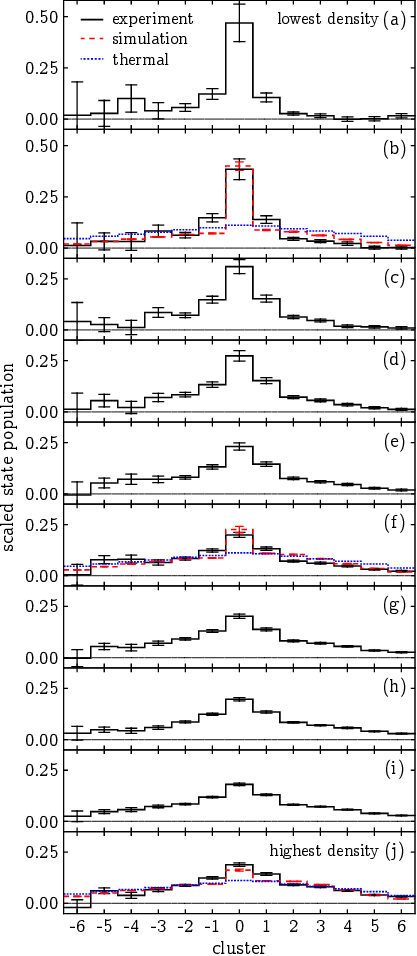}
\caption{\label{fig:density-bar} The scaled and normalized population of each cluster for ten experimental densities from about (a) $2\times10^8$~cm$^{-3}$ to (j) $1.5\times10^{10}$~cm$^{-3}$ (solid black). Simulation results are shown in (b), (f), and (j) for the corresponding densities (dashed red). In each case, the simulations are allowed to equilibrate. The predicted thermal equilibrium state is shown for the same panels (dotted blue). Limited signal for clusters above the initial cluster (-6, -5, -4, -3) results in noise that is amplified by the $36d$ state coupling scaling. Statistical error bars are calculated using variations between laser shots for a given density bin.}
\end{figure}

We vary the experimental density by adjusting the number of trapped atoms, while keeping the Rydberg excitation laser intensity and volume constant. The intensity of our hyperfine pumping laser is adjusted using a rotating a half-wave plate in front of a linear polarizing filter. This varies the number of atoms available for Rydberg excitation while keeping the trap volume roughly constant with a diameter of 0.5~mm. The Rydberg lasers are focused to a waist of 50~$\mu$m, selecting an excitation volume at the center of the trapped atom sample. The filter rotates continuously at 2~RPM while we asynchronously collect microwave spectra at 20 samples per second for several hours. Since our Rydberg excitation volume remains constant, the total Rydberg signal level for each excitation is a reasonable proxy for density. We therefore separate our data at each microwave frequency into ten equal bins according to the total signal level measured on each excitation cycle. The scaled cluster populations for these ten density bins are shown in Fig.~\ref{fig:density-bar}. The error bars associated with the experimental data represent the standard error of the mean of the data points collected at each density bin for each cluster. Using the method described in~\cite{opsahl_EnergyTransportHighlyPolarized_2026}, we roughly estimate the Rydberg atom density to range from about $2\times10^{8}$~cm$^{-3}$ to $1.5\times10^{10}$~cm$^{-3}$, which correspond to interatomic spacings of about 18~$\mu$m to 4~$\mu$m respectively. We are currently developing a technique to more accurately measure the Rydberg atom density.

\textit{Numerical Analysis\textemdash}We simulate the experiment by numerically solving the Schr\"{o}dinger equation using a Hamiltonian that includes two-body dipole-dipole interactions, as in~\cite{opsahl_EnergyTransportHighlyPolarized_2026}. Since the cluster spacings are nearly harmonic and there are multiple energy levels per cluster, the two-body interactions tend to have small detunings on the order of the spacing between energy levels within a cluster. This opens up the possibility of three-body interactions similar to those reported in~\cite{faoro_BorromeanThreebodyFRET_2015,tretyakov_ObservationBorromeanThreeBody_2017,liu_TimeDependenceFewBody_2020}, in which a small intra-cluster spacing could mostly cancel the detuning of the two-body interaction. Thus, we also include three-body dipole-dipole interactions. The model includes 52 Stark energy levels, which are arranged into 13 clusters of four $|m_j|=1/2$ and $|m_j|=3/2$ energy levels each. 

Including only four atoms, ignoring angular dependence, and ignoring states that are detuned by more than the typical inter-cluster spacing, the resulting Hamiltonian has size $376{,}064$. This is too large for exact diagonalization; however, the Hamiltonian is extremely sparse which allows for the efficient implementation of a fourth-order Runge-Kutta method~\cite{elsayed_RegressionRelationPure_2013} using sparse matrix libraries~\cite{anzt_GinkgoModernLinear_2022}. Assuming an initial state that is an equal superposition of all levels in the initially excited cluster, we simulate the time evolution to equilibration. Given the uncertainty in our density calibration and the limits of our model, we do not expect quantitative agreement. However, the simulation results in Fig.~\ref{fig:density-bar} do qualitatively agree with the data, showing that significant population remains in the initial cluster, especially at lower densities.  

Since the dipole-dipole matrix elements connecting different clusters are similar in size and the anharmonicity of the ladder is small, one intuitively expects the thermal state to be a relatively even spread of energy across the clusters. While both the experimental and simulated energy distributions in Fig.~\ref{fig:density-bar} become more evenly spread as the density increases, they remain more sharply peaked in the center around the initial cluster. To quantitatively check whether the atoms thermalize, we calculate the thermal state.

The microcanonical average for an observable $\hat{A}$ is
\begin{equation}
    \langle\hat{A}\rangle_{mc}=\frac{1}{N_S}\sum_{E_n\in S} \bra{\phi_n}\hat{A}\ket{\phi_n},\label{eq:mcavg}
\end{equation}
where the $\ket{\phi_n}$ are the eigenstates, $S=[E_0-\Delta E,E_0+\Delta E]$ defines an energy shell centered on the energy $E_0$ of our initial state, and $N_S$ is the number of eigenstates within this energy shell. In the present case, there is an observable $\hat{A}$ for the population of each cluster, which we measure in our experiment. Unfortunately, Eq.~(\ref{eq:mcavg}) requires calculating the energy eigenstates, which is not feasible given the size of the Hamiltonian. 

To solve this problem we use dynamical quantum typicality, according to which the vast majority of all pure states will have a similar expectation value for some observable at all times, given that they start with similar expectation values at the initial time and provided that the Hilbert space is sufficiently large~\cite{bartsch_DynamicalTypicalityQuantum_2009,binder_ThermodynamicsQuantumRegime_2018}. Those expectation values approach the thermal state predicted by the microcanonical ensemble, provided the system obeys the wETH~\cite{ikeda_FinitesizeScalingAnalysis_2013,reimann_DynamicalTypicalityApproach_2018}. Based on numerical and theoretical work, we expect that our system should, indeed, obey the wETH~\cite{kuwahara_GaussianConcentrationBound_2020,sugimoto_EigenstateThermalizationLongRange_2022}, though we plan to explore this in future work.

Thus, to predict the final thermal state of our system, we start with a random state of the form
\begin{equation}
    \ket{\psi}=\sum_{i=0}^N a_i\ket{\chi_i},
\end{equation}
where $N$ is the size of the Hilbert space, the $\ket{\chi_i}$ are the Stark manifold basis states, and the $a_i$ are quantum amplitudes. The amplitudes have an absolute value randomly chosen from the interval $[0,1)$, a phase randomly chosen from the interval $[0,2\pi)$, and $\ket{\psi}$ is normalized after being generated. 

We use the method described by Steinigeweg, \textit{et al.} to calculate the thermal expectation value of the fractional population of each cluster~\cite{steinigeweg_PushingLimitsEigenstate_2014}. This is accomplished by applying an energy filter operator $e^{-(\hat{H}-E_0)^2/4\Delta E^2}$ to $\ket{\psi}$, which suppresses the contributions of eigenstates outside of this energy shell. The action of this operator is also calculated using a Runge-Kutta scheme, albeit propagating through imaginary time.

Some care must be taken to select an appropriate energy shell. Eigenstates outside of the shell, especially those in the wings of the energy distribution, will not satisfy the assumption of similar expectation values. We use the kernel polynomial method~\cite{weisse_KernelPolynomialMethod_2006} to estimate the energy eigenvalue spectrum, and select a small range around $E_0$ that corresponds to about the middle third of the eigenstates.

We find that averaging over a few dozen instantiations of $\ket{\psi}$ is more than sufficient for convergence. The results are shown in Fig.~\ref{fig:density-bar}, where the predicted thermal distribution is independent of density; at lower densities it would simply take a longer time to equilibrate. The thermal state is, indeed, relatively flat; the decrease in population from the center to the edges is due to the small anharmonicity. The simulation and the experimental data approach the thermal prediction at the highest density, but in both cases excess population remains in the initial cluster.

\textit{Discussion\textemdash}Preliminary numerical evidence suggests that the failure to thermalize in this experiment is due to the presence of quantum many-body scars. The Hamiltonian for the present system shares a feature with the PXP model in which quantum many-body scars were originally reported~\cite{turner_WeakErgodicityBreaking_2018} and with a few-body system in which quantum many-body scars have also been proposed~\cite{spielman_QuantumManybodyScars_2024}; namely, that there are two interactions, one weak and one strong, and that the strong interactions are suppressed for some initial states.

The strong interactions are the couplings between energy levels within a cluster (around 250~$ea_0$), which have dipole moments that are typically an order of magnitude larger than the weaker couplings between clusters (around 60~$ea_0$). An initial state in which only the bottom or top energy level in a cluster was excited, for example, would present no possible dipole-dipole interactions within the cluster. 


This new experimental platform has the potential to yield deeper understanding of the thermodynamics of strongly coupled, isolated quantum systems. By refining our excitation technique, we can further probe the role of quantum many-body scar states on the nonthermal behavior observed in this system. For example, by moving the initial state closer to the edge of the manifold, we can gain insight into edge effects. Additionally, we can excite a superposition of manifold clusters to explore the dynamics of an initial state that more closely resembles a thermal state.  We plan to study these possibilities in future work. 
\begin{acknowledgments}
This work was supported by the National Science Foundation under Grants No.~2011583, No.~2011610, and No.~2427091, and S.E.S. is supported by the National Science Foundation Graduate Research Fellowship under Grant No. 2334429.

This work used the Delta system at the National Center for Supercomputing Applications through allocation PHY250155 from the Advanced Cyberinfrastructure Coordination Ecosystem: Services \& Support (ACCESS) program, which is supported by National Science Foundation grants \#2138259, \#2138286, \#2138307, \#2137603, and \#2138296.
\end{acknowledgments}

\bibliography{stark-dd-density}

\end{document}